\documentclass[english,prl,aps,reprint,floatfix]{revtex4-2}
\usepackage{ae,aecompl}
\usepackage[T1]{fontenc}
\usepackage[latin9]{inputenc}
\usepackage{color}
\usepackage{amsbsy}
\usepackage{amstext}
\usepackage{graphicx}
\usepackage{babel}
\usepackage[normalem]{ulem}

\begin{document}
\title{System Size Dependence of Collisionless Reconnection Rate}
\author{Yi-Min Huang}
\email{yopology@umd.edu}

\affiliation{Department of Astronomy, University of Maryland, College Park, MD
20742, USA}
\affiliation{National Aeronautics and Space Administration Goddard Space Flight
Center, Greenbelt, MD 20771, USA}
\author{Naoki Bessho}
\affiliation{Department of Astronomy, University of Maryland, College Park, MD
20742, USA}
\affiliation{National Aeronautics and Space Administration Goddard Space Flight
Center, Greenbelt, MD 20771, USA}
\author{Li-Jen Chen}
\affiliation{National Aeronautics and Space Administration Goddard Space Flight
Center, Greenbelt, MD 20771, USA}
\author{Judith T. Karpen}
\affiliation{National Aeronautics and Space Administration Goddard Space Flight
Center, Greenbelt, MD 20771, USA}
\author{Amitava Bhattacharjee}
\affiliation{Department of Astrophysical Sciences, Princeton University, Princeton,
New Jersey 08543, USA}
\begin{abstract}
It is a widely accepted paradigm that collisionless magnetic reconnection proceeds at a universal fast rate of $\sim0.1$ when normalized to a properly defined reconnecting magnetic field and Alfv\'en speed, effectively independent of the macroscopic system size. This conclusion, derived primarily from kinetic simulations of classical Harris current sheets with kinetic-scale thickness, stands in contrast to results from forced reconnection and island coalescence, where the rate significantly depends on the system size. Here, we reconcile this disparity by performing a rigorous scaling study of Harris sheets using both particle-in-cell and Hall magnetohydrodynamic simulations. We demonstrate that when the global magnetic configuration is preserved by scaling the initial current sheet thickness proportionally with the system size, the ``universal'' fast rate disappears. Instead, the reconnection rate decreases as the system size increases. By establishing a consistent scaling trend across different magnetic configurations, these findings challenge the paradigm of scale-independent fast reconnection and necessitate a reevaluation of energy release rates in astrophysical plasmas. 
\end{abstract}
\maketitle
Magnetic reconnection is a fundamental process in laboratory, space,
and astrophysical plasmas, capable of explosively converting magnetic
energy into plasma kinetic energy through changing magnetic field
line topology \citep{PriestF2000,ZweibelY2009,YamadaKJ2010,ZweibelY2016,JiDJLSY2022}.
In many astrophysical environments, such as the solar corona and planetary
magnetospheres, this process occurs in regimes where collisional effects
are weak or negligible. In planetary magnetospheres, collisions are
negligible due to the low plasma density. In the solar corona, even
if the onset of reconnection may occur at magnetohydrodynamic (MHD)
scales, the fragmentation of reconnection current sheets likely will
cascade to kinetic scales, triggering transition to collisionless
reconnection \citep{DaughtonRAKYB2009,HuangBS2011,JiD2011}. Therefore,
modeling reconnection in these astrophysical environments requires
the inclusion of collisionless physics. At the fundamental level,
collisionless reconnection is often modeled using fully kinetic particle-in-cell
(PIC) simulations. However, the immense scale separation in astrophysical
systems forces computational compromises; while the ion to electron
mass ratio and the speed of light are often reduced, the disparity in system size is perhaps the most staggering. For instance, typical
solar flares \citep{KumarKAWD2019} and reconnection in coronal mass
ejection (CME) current sheets \citep{GuoBH2013,LinMSRRZWL2015} and
heliospheric current sheets \citep{PhanBE2020,PhanDL2024} operate
on a macroscopic scale $L\sim10^{7}-10^{9}d_{i}$ (ion skin depth),
a regime many orders of magnitude beyond current kinetic simulation
capabilities. Consequently, determining how reconnection scales with system size is not merely a numerical exercise but a prerequisite
for extrapolating simulation results to reality.

Within the realm of fundamental research on magnetic reconnection
physics, the classical Harris current sheet \citep{Harris1962} serves as the
\emph{de facto} initial condition for the majority of studies. Based
on a seminal scaling study from the turn of the century using the
Harris sheet configuration \citep{ShayDRD1999}, a paradigm has been
established that collisionless reconnection proceeds at a ``universal''
fast rate of approximately $0.1$ when normalized to a properly defined
reconnecting magnetic field and Alfv\'en speed, effectively independent
of the system size. This conclusion has become widely accepted in
the community, reinforced by the ubiquity of the $\sim0.1$ rate in standard kinetic Harris sheet simulations \citep{HubaR2004,CassakLS2017,LiuHGDLCS2017}.
Because simulations of varying sizes typically yield rates within
a factor of unity of this value, size independence
has been adopted as a standard working assumption and remained largely unquestioned.

However, this accepted wisdom stands in contrast to results from alternative
configurations such as forced reconnection \citep{WangBM2001,Fitzpatrick2004} and magnetic island coalescence \citep{KarimabadiDRDC2011,StanierDCKNHHB2015,NgHHBSDWG2015}, which
consistently report a significant dependence on the system size. In these
configurations, the reconnection rate is found to decrease as the
system scale $L$ increases relative to the kinetic scale $d_{i}$.
This disparity between the standard Harris sheet results (size-independent)
and alternative configurations (size-dependent) is typically attributed
to differences in global magnetic topology or boundary forcing. 

We propose that this disparity stems instead from a fundamental difference
in how the system size scaling studies are performed. Standard Harris
sheet simulations typically initialize the current sheet with a thickness
$\delta$ on the order of $d_{i}$. This choice is motivated by computational
efficiency, bypassing the long phase required for a thick sheet to
thin down to kinetic scales to trigger the onset of collisionless
reconnection. However, when investigating system size scaling, these
studies almost invariably increase the simulation domain size $L$
(often only along the outflow direction, e.g., Refs.~\citep{ShayDRD1999,DaughtonRAKYB2009})
while keeping the initial sheet thickness $\delta$ fixed at the kinetic
scale. As a result, the ratio $\delta/L$ varies, effectively changing
the global magnetic configuration as the system size increases. A
rigorous study of system size scaling requires preserving the global
magnetic configuration while scanning the scale separation $L/d_{i}$
as an asymptotic parameter. This implies that as the system size $L$
increases, the macroscopic features---including the initial current
sheet thickness---must scale proportionally. Furthermore, to compare
reconnection rates across different system sizes, the rates must be
normalized to some fixed reference Alfv\'en speed and magnetic field
associated with the global configuration, rather than to dynamically
measured local upstream values because the latter may depend on the
system size and thereby mask the true system size dependence. These
rigorous scaling protocols are abided by the forced reconnection and island coalescence studies that report size dependence. In this Letter,
we demonstrate that when the same protocols are applied to the standard
Harris sheet, the universal, size-independent rate disappears. We
find that the reconnection rate decreases as the system size increases,
thereby demonstrating that dependence on macroscopic scales is not peculiar to specific geometries like island coalescence, but is an intrinsic property shared by the standard Harris sheet.

To illustrate the point, we employ the Geospace Environmental Modeling (GEM) challenge problem \citep{BirnDSRDHKMBOP2001} as an example. Let us consider a Harris sheet equilibrium with a magnetic field 
\begin{equation}
B_{x}\left(z\right)=B_{0}\tanh\left(z/\delta\right)\label{eq:harris}
\end{equation}
and a density profile 
\begin{equation}
n\left(z\right)=n_{0}\text{sech}^2\left(z/\delta\right)+n_{\infty}.\label{eq:density}
\end{equation}
The electron and ion temperatures, $T_{e}$ and $T_{i}$, are taken
to be uniform in the initial state. Therefore, the pressure balance
condition requires $n_{0}k_{B}\left(T_{e}+T_{i}\right)=B_{0}^{2}/2\mu_{0}$
(in SI units). The two-dimensional (2D) system is in a rectangular
domain with $-L_{x}/2\le x\le L_{x}/2$ and $-L_{z}/2\le z\le L_{z}/2$.
Periodic and perfectly conducting boundary conditions are imposed
in the $x$ and $z$ directions, respectively. 

The GEM challenge problem specifies the density and the temperature
ratios to be $n_{\infty}/n_{0}=0.2$ and $T_{e}/T_{i}=0.2$. The ion
to electron mass ratio is $m_{i}/m_{e}=25$. Using the ion skin depth
$d_{i}\equiv\sqrt{m_{i}/\mu_{0}n_{0}e^{2}}$ as the reference length
$\ell$, the domain sizes $L_{x}=25.6\ell$ and $L_{z}=12.8\ell$;
the initial Harris sheet half width $\delta=0.5\ell$. Reconnection
is initiated by adding a perturbation $\tilde{\boldsymbol{B}}=\boldsymbol{\hat{y}}\times\nabla\tilde{\psi}$
to the magnetic field, where 
\begin{equation}
\tilde{\psi}=0.1B_{0}\ell\cos\left(2\pi x/L_{x}\right)\cos\left(\pi z/L_{z}\right).\label{eq:psi_perturb}
\end{equation}

To scale the GEM challenge problem to large system sizes, we vary
$\ell/d_{i}$ as a free parameter, equivalent to changing the system
size while keeping macroscopic ratios, therefore the global configuration,
fixed. A key conclusion of the GEM challenge is that Hall MHD is a
minimal model that captures the essence of collisionless reconnection.
We perform both fully kinetic PIC simulations using the WarpX code
\citep{zenodo.15604100} and Hall MHD simulations using the HMHD code
\citep{HuangBS2011,HuangB2024}. This Letter reports a scaling study
that varies $\ell/d_{i}$ from unity to $30$ $(100)$ for PIC (Hall
MHD) simulations. The largest case with $\ell/d_{i}=30$ $(100)$
corresponds to a system size of $768d_{i}\times384d_{i}$ ($2560d_{i}\times1280d_{i}$)
and an initial Harris sheet half width of $15d_{i}$ ($50d_{i}$).
Importantly, we present our results from a global perspective: time
is normalized to $t_{A}=\ell/V_{A}$ and reconnection rate is
normalized to $B_{0}V_{A}$, where $V_{A}=B_{0}/\sqrt{\mu_{0}n_{0}m_{i}}$.
For the GEM challenge system size with $\ell/d_{i}=1$, the time unit
$t_{A}$ becomes the usual ``microscopic'' choice for the time unit---the
reciprocal of the ion cyclotron frequency $1/\omega_{ci}$, making
it easy to compare with GEM challenge literature. 

For PIC simulations, the ratio of electron plasma frequency to cyclotron
frequency $\omega_{pe}/\omega_{ce}=2$; the spatial resolution is
20 grid cells per $d_{i}$ in each direction, with an average of 250
macro particles per cell for each species. For Hall MHD simulations,
we employ hyper-resistivity to break the frozen-in condition \citep{HuangBF2013}.
We set the hyper-resistivity coefficient $\eta_{H}$ such that the
hyper-resistive Lundquist number based on the length scale $d_{i}$
is fixed at $d_{i}^{3}V_{A}/\eta_{H}=10^{4}$. Likewise, we set the
ion viscosity $\nu$ such that $d_{i}V_{A}/\nu=10^{2}$. These choices
ensure that dissipative effects remain the same at the $d_{i}$ scale
across different simulations. Electron and ion temperatures are stepped
separately with their respective velocities and an adiabatic equation
of state. We employ a nonuniform grid for Hall MHD simulations, with
grid points packed near the origin to resolve the dissipation region surrounding
the dominant X-point. In the vicinity of the origin, the grid spacing
$\Delta x\simeq0.1d_{i}$ along the $x$ direction and $\Delta z\simeq0.02d_{i}$
along the $z$ direction. We further assume an up-down symmetry and
only simulate the region $z\ge0$. 

\begin{figure}[htbp]
\includegraphics[width=0.9\columnwidth]{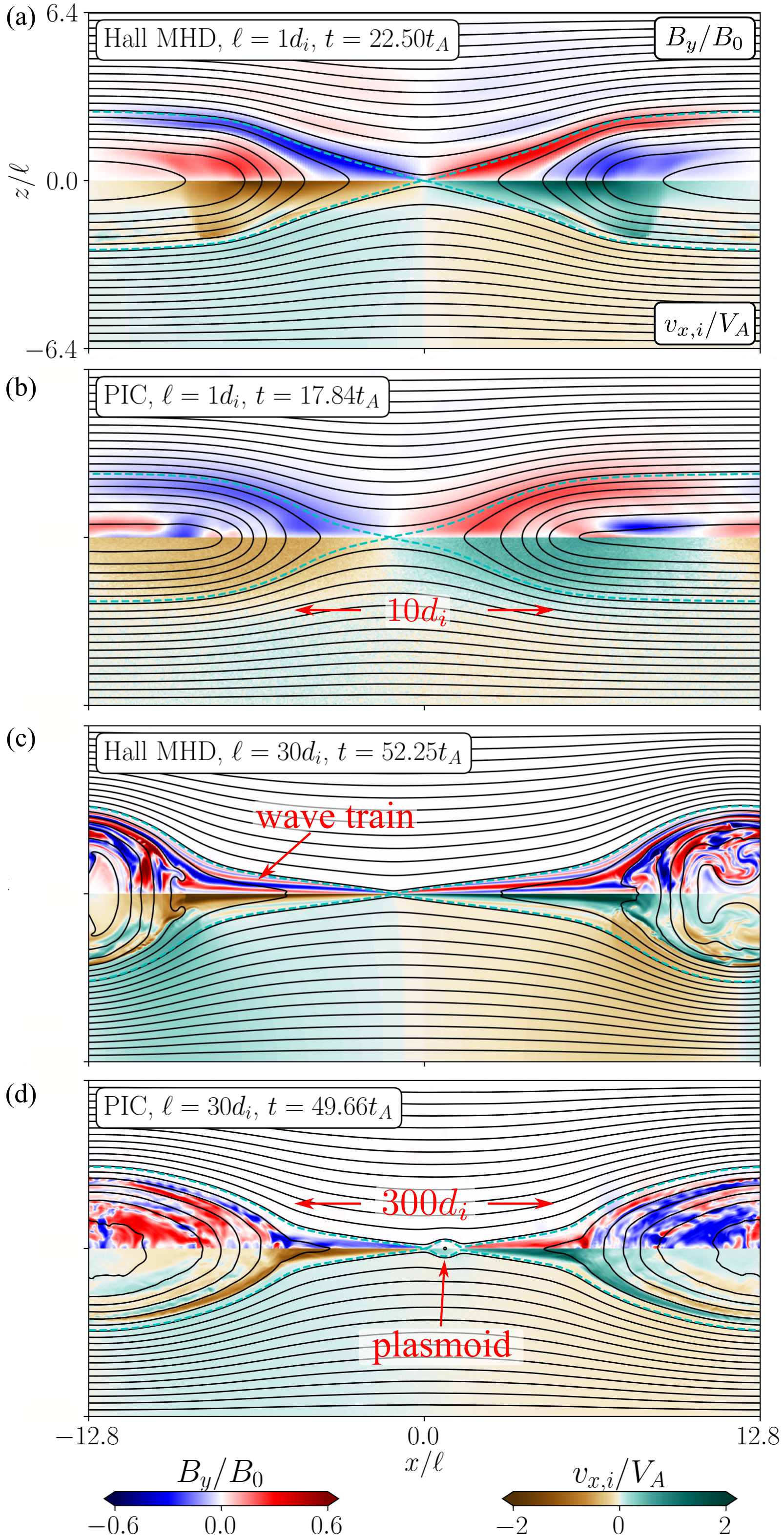}

\caption{Snapshots of Hall MHD (a: $\ell=d_{i}$, c: $\ell=30d_{i}$) and PIC
(b: $\ell=d_{i}$, d: $\ell=30d_{i}$) simulations when $\psi_{\text{rec}}/B_{o}\ell\approx2.0$.
The upper half shows the normalized out-of-plane magnetic field $B_{y}/B_{0}$,
which is a signature of the Hall effect. The lower half shows the
normalized ion outflow velocity $v_{x,i}/V_{A}$. Black solid lines
are magnetic field lines, and cyan dashed lines are separatrices that
divide reconnected and unreconnected regions. \protect\label{fig:4-panels}}
\end{figure}

\begin{figure}[htbp]
\includegraphics[width=0.9\columnwidth]{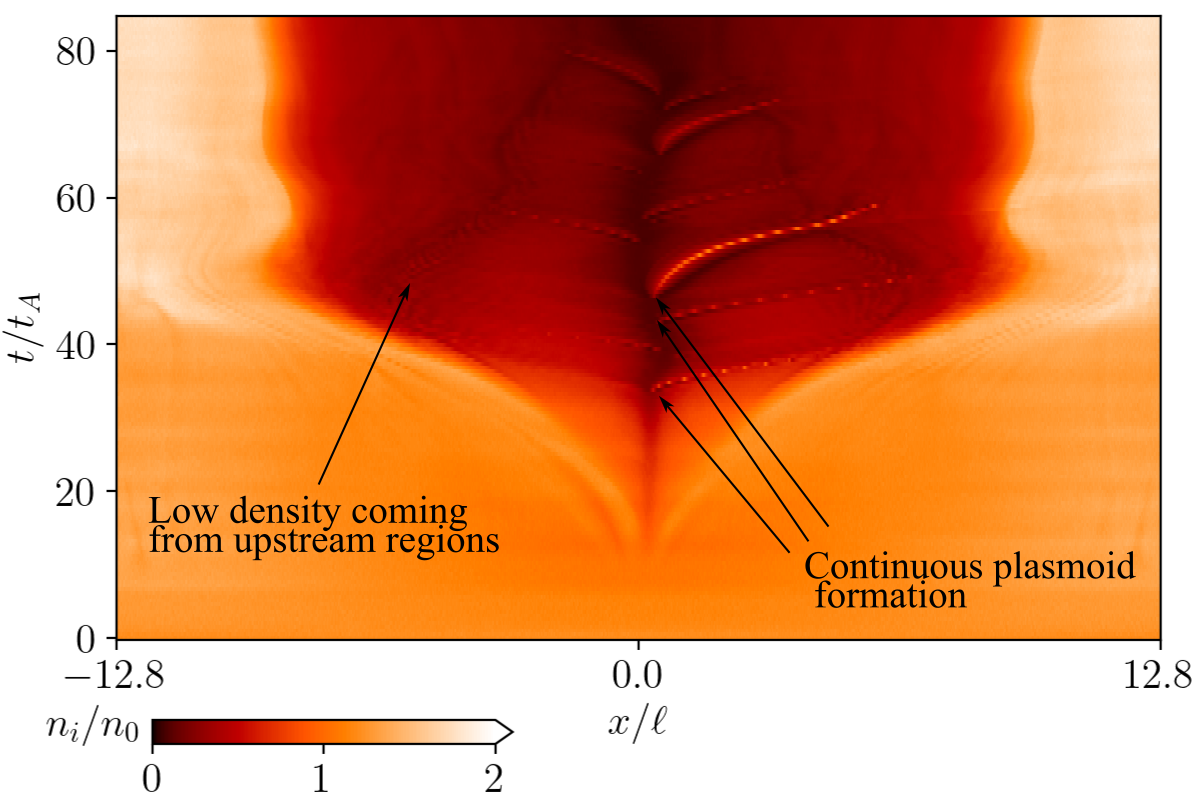}

\caption{Spacetime plot of the ion density along the mid-line ($z=0$) for
the PIC simulation with $\ell=30d_{i}$. \protect\label{fig:density-stack}}
\end{figure}

The Harris sheet in all PIC simulations becomes unstable as it thins
down, leading to formation of plasmoids (see discussion below and
Figures \ref{fig:4-panels} and \ref{fig:density-stack}). We also
seed a random noise of amplitude $10^{-3}V_{A}$ to the initial ion
velocity in Hall MHD simulations to break the left-right symmetry,
preventing the undesirable situation of a plasmoid sitting at the
center of the current sheet and throttling reconnection. For this
set of Hall MHD simulations, plasmoids are observed in cases with $\ell/d_i\ge3$.
In all cases, the onset of Hall reconnection expels all the plasmoids
and the system settles to a single-X-point geometry. Within the Hall MHD
model, single-X-point geometry tends to yield a faster rate than the
intermediate regime where the single-X-point geometry is unstable and new plasmoids continue to form \citep{HuangBS2011}. 

\begin{figure}[htbp]
\includegraphics[width=0.9\columnwidth]{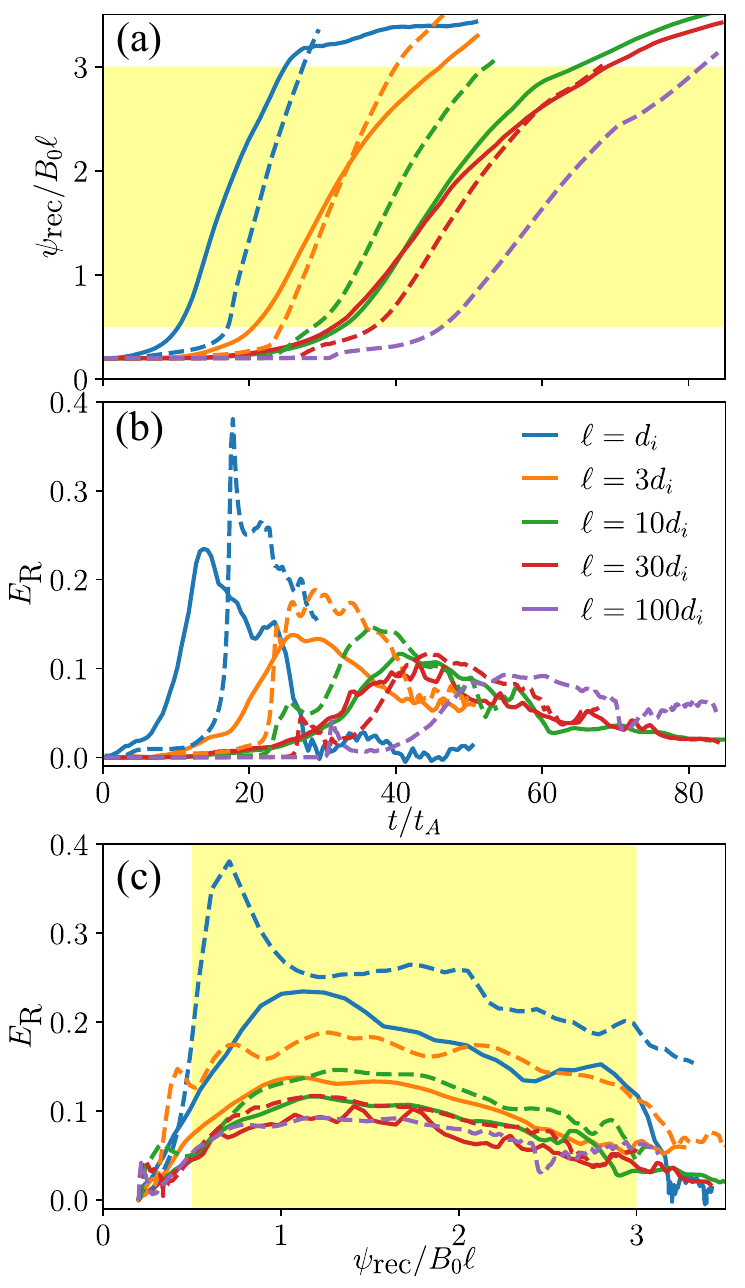}

\caption{(a) Normalized reconnected flux $\psi_{\text{rec}}/B_{0}\ell$ and
(b) normalized reconnection rate versus normalized time $t/t_{A}$;
(c) normalized reconnection rate versus normalized reconnected flux.
Line colors correspond to different system sizes; solid lines are
PIC simulations and dashed lines are Hall MHD simulations. Yellow
shaded regions denote the interval of reconnected flux that will be
used to evaluate the averaged reconnection rate. \protect\label{fig:Reconnection_rate}}
\end{figure}

We measure the reconnected flux $\psi_{\text{rec}}$ by first inverting
$\boldsymbol{\hat{y}\times\nabla\psi}=B_x\boldsymbol{\hat{x}}+B_z\boldsymbol{\hat{z}}$ to obtain the
flux function $\psi$ for PIC simulations (the 2D version of HMHD
uses $\psi$ as a primary variable) then calculating $\psi_{\text{rec}}=\max\left.\psi\right|_{z=0}-\text{\ensuremath{\min}}\left.\psi\right|_{z=0}$.
The time derivative of $\psi_{\text{rec}}$ yields the reconnection
rate. Figures \ref{fig:Reconnection_rate}(a) and (b) show
the normalized reconnected flux $\psi_{\text{rec}}/B_{0}\ell$ and
the normalized reconnection rate $E_{R}\equiv\left(d\psi_{\text{rec}}/dt\right)/V_{A}B_{0}$
versus the normalized time $t/t_{A}$ for PIC and Hall MHD simulations
with varying system sizes. The PIC and Hall simulation results overall
follow the same trend. As the system size becomes larger, the onset
of reconnection occurs later due to the longer initial thinning time
of a thick current sheet toward kinetic scales. To compare the reconnection
rates across different runs, we plot the normalized reconnection rate
against the normalized reconnected flux, shown in Figure \ref{fig:Reconnection_rate}(c).
Because the reconnected flux serves as a proxy for the global configuration, this plot allows us to compare the reconnection rates of different runs under similar macroscopic conditions. The results shown
in Figure \ref{fig:Reconnection_rate}(c) effectively align the periods of substantial reconnection rates for all cases into the same range of reconnected flux. Two features are immediately noticeable. First, the reconnection rates in Hall MHD runs are consistently slightly higher than in the corresponding PIC runs. Second, and more importantly, as the system size increases, the reconnection rate decreases. 

To demonstrate the similarity of global configurations across different
runs when the reconnected fluxes are the same, Figure \ref{fig:4-panels}
shows snapshots of Hall MHD and PIC simulations for two different
system sizes with $\ell=d_{i}$ and $\ell=30d_{i}$ and the reconnected
flux $\psi_{\text{rec}}/B_{o}\ell\approx2.0$. In each panel, color
coding shows the out-of-plane magnetic field $B_{y}$ in the upper
half and the outflow component of the ion velocity $v_{x,i}$ in the
lower half (both are normalized). Since the reconnected fluxes are
approximately the same for these snapshots, the global configurations,
such as the sizes of the main magnetic islands at the $x$ boundaries, are similar. However, the opening
angles of the separatrices in the exhaust regions are considerably
narrower for the $\ell=30d_{i}$ cases than the $\ell=d_{i}$ cases.
This trend of a decreasing opening angle as the system size increases
is consistently observed in both Hall MHD and PIC simulations. Because
the opening angle of the exhaust region is related to the reconnection
rate \citep{LiuCLHLG2022}, this trend is consistent with the result
that the reconnection rate decreases as the system size increases. This narrowing is a physical manifestation of how the system retains a ``memory'' of its macroscopic scale. The separatrix opening angle is ultimately determined by a competition between the localized kinetic processes attempting to open the exhaust and the global macroscopic magnetic stress resisting it. As the system size $L$ increases, the diffusion region at the $d_i$ scale becomes increasingly restricted by the massive macroscopic stress, forcing the exhaust into a narrower angle and limiting the global reconnection rate.

The overall global configurations between PIC and Hall MHD simulations
at the same system size are similar, but there are some noticeable
differences. The outflow jets are narrower and more collimated in
Hall MHD simulations than in PIC simulations. At large system sizes,
Hall MHD solutions exhibit wave trains in both $B_{y}$ and $v_{x,i}$
variables in regions between the exhaust jets and the separatrices
(Figure \ref{fig:4-panels}(c)). These wave trains are likely a signature
of dispersive shocks due to whistler waves \citep{HauW2016}, but PIC simulations do not exhibit these features. One potential reason for this discrepancy is that the continuous generation and ejection of plasmoids in PIC simulations could disrupt the coherence of these wave trains, though further investigation into whistler wave dispersion and damping in the kinetic regime is required. As the system size becomes larger, the number of plasmoids throughout
a PIC simulation increases. The continuous formation of plasmoids
is clearly visible in Figure \ref{fig:density-stack}, showing the
ion density spacetime plot along the mid-line ($z=0$) for the case
with $\ell=30d_{i}$. In this Figure, plasmoids are high-density streaks
that originate near the center and propagate toward the downstream
regions with the reconnection outflow jets. The snapshot in Figure
\ref{fig:4-panels}(d) shows a plasmoid near the center of the domain. 

\begin{figure}[htbp]
\includegraphics[width=0.9\columnwidth]{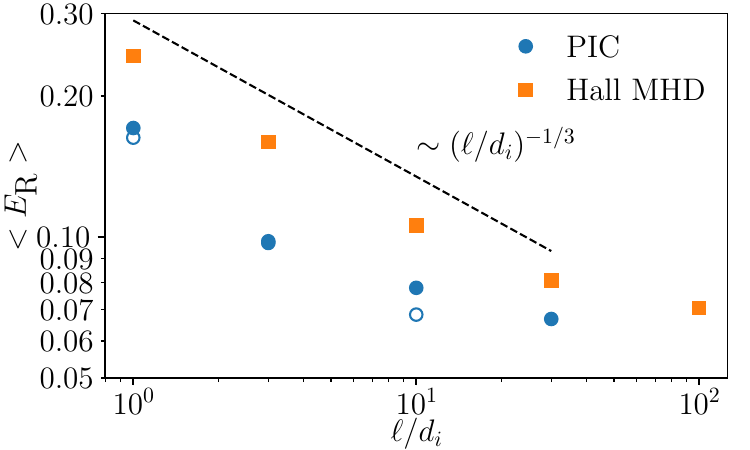}

\caption{Scaling of average reconnection rates for PIC (solid blue circles)
and Hall MHD (solid orange squares) simulations in a log-log plot.
Empty blue circles denote additional PIC runs for $\ell=d_{i}$,
$3d_{i}$, and $10d_{i}$. The average rate of the additional run
for $\ell=3d_{i}$ is nearly indistinguishable from that of the reference
run. \protect\label{fig:Scaling-of-average-reconnection=000020rate}}
\end{figure}

We measure the average reconnection rate during the period when $0.5\le\psi_{\text{rec}}/B_{0}\ell\le 3$,
corresponding to the plateau of reconnection rates highlighted as
the yellow shaded regions in Figure \ref{fig:Reconnection_rate}.
The average reconnection rate $\left\langle E_{R}\right\rangle $$=\left(\Delta\psi_{\text{rec}}/\Delta t\right)/V_{A}B_{0}$,
where $\Delta t$ is the time it takes for the reconnected flux to
increase from $\psi_{\text{rec}}=0.5B_{0}\ell$ to $3B_{0}\ell$.
The scaling of average reconnection rates $\left\langle E_{R}\right\rangle $
with respect to $\ell/d_{i}$, for PIC (solid blue circles) and Hall
MHD (solid orange squares) simulations, are shown in Figure \ref{fig:Scaling-of-average-reconnection=000020rate}.
The average reconnection rates of Hall MHD simulations approximately
follow a $\left\langle E_{R}\right\rangle \sim\left(\ell/d_{i}\right)^{-1/3}$
scaling law when $\ell/d_{i}<30$, and the scaling becomes less steep
for $\ell/d_{i}\ge30$. In contrast, the scaling of reconnection
rates from PIC simulations is less definitive. Because the initial particle loading randomly varies between independent runs, the stochastic nature of plasmoid formation causes the average reconnection rate to vary even under identical macroscopic parameters. We assess these variations by performing an additional
set of PIC simulations for the three cases with $\ell=d_{i}$, $3d_{i}$,
and $10d_{i}$. The results of these additional runs are shown in
Figure \ref{fig:Scaling-of-average-reconnection=000020rate} with
empty blue circles. For the first two cases, the average rates from
additional runs are nearly identical to those of the reference runs;
whereas for the $\ell=10d_{i}$ case, the additional run is approximately
$12\%$ slower than the reference run due to a large plasmoid lingering
near the center of the system for an extended time before being ejected. This noticeable variance between the two $\ell=10 d_{i}$ runs is a direct consequence of the highly bursty and unsteady dynamics introduced once the extended current sheet becomes unstable to secondary tearing.
Despite the uncertainties of PIC reconnection rates due to plasmoid
formation, the average PIC reconnection rates exhibit a trend similar
to the one observed in Hall MHD, namely, the scaling is steeper at smaller sizes
and becomes more gradual as the system size becomes larger. Within the range
of system sizes scanned, the PIC reconnection rate decreases by a
factor of approximately $2.6$ when the system size increases by a factor of 30. Similarly, the Hall MHD reconnection rate decreases by a factor of approximately $3.4$
when the system size increases by a factor of 100. It remains an open
question whether the decrease of reconnection rate will continue to
larger system sizes or will eventually plateau. 

In conclusion, this study highlights the importance of adhering to rigorous protocols when performing a system size dependence study. When the protocols are strictly followed, both PIC and Hall MHD simulations demonstrate size-dependent reconnection rates with a Harris sheet setup. The fact that Hall MHD and PIC simulations represent two extremes in the spectrum of modeling collisionless reconnection, with Hall MHD being barely minimal and PIC being first-principles, indicates that system size dependence of the collisionless reconnection rate is robust. Taken together with previous findings from 2D forced reconnection and island coalescence studies, a consistent trend emerges across different global configurations: the collisionless reconnection rate declines with increasing system size. It is intriguing to note that existing observational evidence aligns with this qualitative trend of the reconnection rate declining with the system size. \textit{In situ} measurements by the Magnetospheric Multiscale (MMS) mission in the Earth's relatively small magnetosphere ($L\sim 10^3d_i$) consistently reveal fast normalized reconnection rates of 0.05 to 0.3 \citep{GenestretiNND2018, NakamuraGLN2018, ChenWHE2019}. In contrast, the normalized rates inferred from remote observations of solar flares ($L\sim 10^7-10^9d_i$) are often significantly slower, in the range of 0.001 to 0.2 \citep{OhyamaS1998, YokoyamaAMIN2001, IsobeYSMKENS2002, QiuLGW2002, IsobeTS2005,  LinKSRSJZM2005}. While these disparate astrophysical environments differ in many fundamental physical parameters beyond just system size, this massive scale disparity offers a compelling motivation for rigorously investigating the size dependence of the reconnection rate.

This conclusion does not imply that there is a universal system size scaling of the reconnection rate across different configurations. Indeed, previous studies report a diverse range of dependencies sensitive to both the global topology and the chosen kinetic model or fluid closure \citep{WangBM2001,Fitzpatrick2004,KarimabadiDRDC2011,StanierDCKNHHB2015,NgHHBSDWG2015}.  This variability highlights the importance of understanding the essential physics dictating size-dependencies---such as why Hall MHD approximates Harris sheet dynamics relatively well, yet predicts a significantly weaker dependence than PIC simulation does for reconnection rates in island coalescence \citep{StanierDCKNHHB2015,NgHHBSDWG2015}. Moreover, size-dependencies can evolve as the scale separation ($L/d_i$) increases; even within the present study, the decline in the reconnection rate becomes more gradual at larger system sizes.

While this study establishes the system-size dependence of the collisionless reconnection rate for Harris sheets, several physical complexities warrant further investigation. For instance, our PIC simulations employ a reduced ion-to-electron mass ratio $m_{i}/m_{e}=25$.
Although some previous studies suggest that the reconnection rate's dependence on the mass ratio may be weak \citep{HesseSBK1999, RicciLB2002a}, this conclusion requires further verification in the present context of system-size scaling. Furthermore, our PIC simulations indicate that plasmoid formation becomes more prominent as the system size increases. Whether these plasmoids ultimately prevent the collisionless reconnection rate in large systems from further declining remains a topic of debate \citep{DaughtonSK2006, ShayDS2007}. Moreover, the role of boundary conditions warrants future investigation. While our periodic domain retains the system's ``memory'' of its initial configuration, studies utilizing open boundary conditions \citep{HubaR2004, DaughtonSK2006} allow initial plasma to escape, potentially leading to different scaling behaviors. Open boundary conditions were originally devised to mimic systems much larger than the actual simulation domain and eliminate the quenching effect of the boundary in the exhaust region. However, we emphasize that as long as reconnection cannot deplete all the available flux within one global Alfv\'en time, quenching from macroscopic boundaries is a physical reality of finite systems rather than an artifact to be avoided.  Finally, the effects of a guide field, dynamic current sheet formation, and three-dimensional (3D) turbulence \cite{DaughtonRKYABB2011,HuangB2016,StanierDLLB2019, LeakeDK2020, DaldorffLK2022, HuangB2024} must eventually be incorporated. Nevertheless, we note that existing PIC studies have found no dramatic difference between 2D and 3D collisionless reconnection rates under a number of relevant conditions, suggesting that the 2D scaling presented here provides a foundational baseline. Resolving these questions through theoretical development and numerical simulations is an important step toward understanding reconnection in space and astrophysical plasmas. Since current computational domains are orders of magnitude smaller than large astrophysical systems, directly extrapolating these scaling trends remains speculative. However, if the observed declining trend continues to larger system sizes, collisionless reconnection in astrophysical events would be significantly slower than previously expected. Unraveling the macroscopic scaling of collisionless reconnection is therefore an essential next step for accurately modeling explosive energy release in astrophysical plasmas.

\textit{Acknowledgments}---Y.-M.H. dedicates this paper to the memory of Professor Adil Hassam, whose enduring mentorship and physical insights were a profound inspiration for this work. This research was supported by the National Science Foundation, grant number AGS-2301337, the National Aeronautics and Space Administration, grant number 80NSSC25K0050, and the Department of Energy, grant number DE-SC0026063. Computations were performed on facilities at the National Energy Research Scientific Computing Center. This research used the open-source particle-in-cell code WarpX. Primary WarpX contributors are with LBNL, LLNL, CEA-LIDYL, SLAC, DESY, CERN, Helion Energy, and TAE Technologies. We acknowledge all WarpX contributors.

\textit{Data availability}---The data that support the findings of this study are openly available in Zenodo at https://doi.org/10.5281/zenodo.23875157, Ref.~\cite{Huang2026}.

\bibliographystyle{apsrev4-2}
\bibliography{ref}

\end{document}